# SVS-net: A Novel Semantic Segmentation Network in Optical Coherence Tomography Angiography Images

Yih-Cherng Lee,[1,#] Ling Yeung,[2,3,#]

[1] Department of Electrical Engineering, National Taiwan University, Taiwan
[2] Department of Ophthalmology, Chang Gung Memorial Hospital, Keelung, Taiwan.
[3] College of Medicine, Chang Gung University, Taoyuan, Taiwan.
[#] Both authors contributed equally to this work.

*Abstract-* **Automated vascular segmentation on optical coherence tomography angiography (OCTA) is important for the quantitative analyses of retinal microvasculature in neuroretinal and systemic diseases. Despite recent improvements, artifacts continue to pose challenges in segmentation. Our study focused on removing the speckle noise artifact from OCTA images when performing segmentation. Speckle noise is common in OCTA and is particularly prominent over large non-perfusion areas. It may interfere with the proper assessment of retinal vasculature. In this study, we proposed a novel Supervision Vessel Segmentation network (SVS-net) to detect vessels of different sizes. The SVS-net includes a new attention-based module to describe vessel positions and facilitate the understanding of the network learning process. The model is efficient and explainable and could be utilized to reduce the need for manual labeling. Our SVS-net had better performance in accuracy, recall, F1 score, and Kappa score when compared to other well recognized models.**

## Introduction

The quantification of retinal microvasculature is important for the diagnosis, classification and prediction of treatment outcomes in neuroretinal and macular diseases. [1-4] Furthermore, changes in retinal microvasculature can also be useful biomarkers for many systemic comorbidities such as hypertension, cardiovascular diseases, neurological diseases, renal diseases, etc. [5-9]

Retinal vessels of different sizes (from arteriole / venule to capillary) can now be detected by optical coherence tomography angiography (OCTA) - a rapid and non-invasive method. [10] Quantitative microvascular analysis within the central area of retina, delineated by Early Treatment Diabetic Retinopathy Study (ETDRS) circles, **Figure 1(a-b)**, is particularly useful. Vessel densities, non-perfusion area sizes and other quantitative parameters within the parafoveal area have been commonly used in recent studies. [1-3,6] Although the microvascular structures are obtained through 3D volume rendering, they are usually presented as 2D images for clinical interpretation. For example, the OCTA machine (Optovue RTVue XR Avanti; Optovue Inc., Fremont, CA) used in this study automatically segments the retinal microvasculature into four layers: superficial vascular plexus, deep vascular plexus, outer retina, and choriocapillary, **Figure 1(c-f)**.

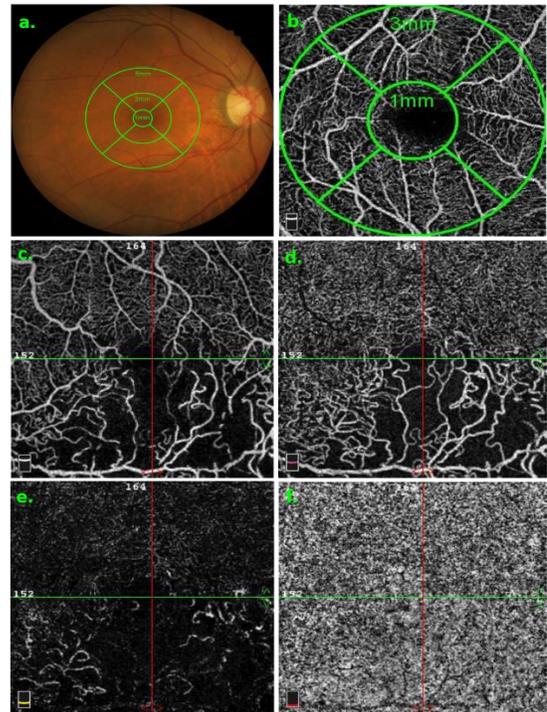

**Figure 1.** Illustration of retinal vessels. **(a)** A color fundus photo of a healthy subject. The diameters of ETDRS circles are 1mm (central), 3 mm (inner) and 6mm (outer), respectively. **(b)** A 3×3 mm² OCTA image of the eye in (a). Parafovea area is the annular area between 1mm- and 3mm-circles. **(c-f)** OCTA images of an eye with branch retinal vein occlusion. The retinal vasculatures were automatically segmented into 4 layers by the machine software. **(c)** Superficial vascular plexus. **(d)** Deep vascular plexus. **(e)** Outer retina. **(f)** Choriocapillaris.

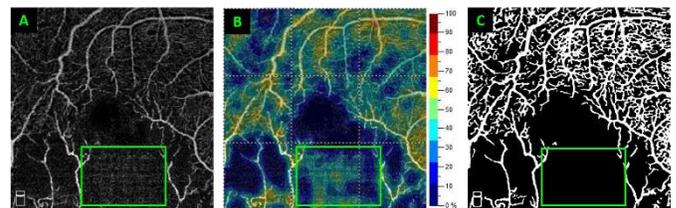

**Figure 2.** The speckle noise within a non-profusion area, as denoted by the green rectangular block, is misclassified as vessels rather than as an avascular region. **(a)** The original OCTA image. **(b)** Vessel density map calculated by the OCTA machine. **(c)** Vascular map produced from our SVS-net model.

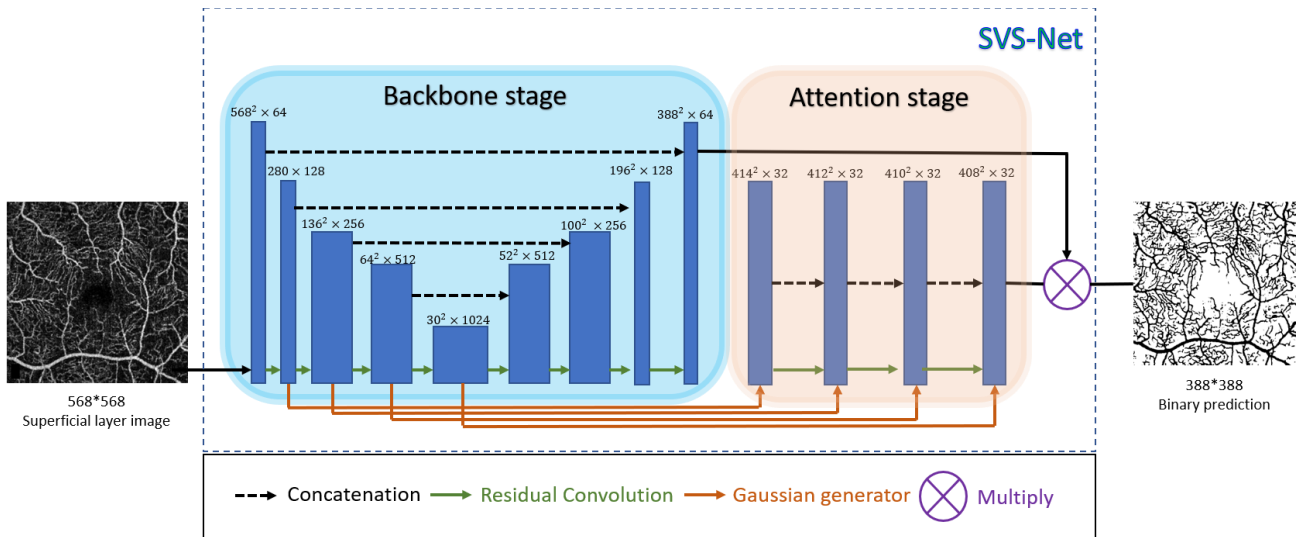

**Figure 3.** The proposed architecture of encoder **(a)** and decoder **(b)** neural network in this study.

Although the OCTA machine built-in software provides acceptable vascular information across retinal layers, it performs poorly when presented with noise artifacts in eyes that contain large non-perfusion areas, as shown in **Figure 2(a)**. The colors in **Figure 2(b)** denote the vessel densities calculated by the OCTA machine software. The diffuse speckle noise over the non-perfusion area had been misclassified as vessels and, therefore, the machine overestimated vessel density within this area.

Several studies have explored machine learning approaches to vessel segmentation in medical images. The approaches include conventional machine learning based methods [11-12], which used various thresholding and filtering techniques; and deep learning-based methods [13-17]. Li et al. [18] and Zhang et al. [19] introduced a 3D image network that can segment 3D vessel images. However, these models have not been evaluated for their effectiveness in the removal of diffuse speckle noise in eyes with large non-perfusion areas.

In this study, we proposed a novel Supervision Vessel Segmentation network (SVS-net) to detect retinal vessels of different sizes in OCTA. The model was trained and tested with an OCTA image database containing images of both normal eyes and eyes with varying non-perfusion area sizes and speckle noise amounts. Our main contributions include:

**(I)** compiled a database with 60 OCTA images of superficial vascular plexus - each with a different level of non-perfusion and precise annotations.

**(II)** designed an explainable feature extraction method. We generated an attention stage to help enhance the clarity and the transparency of the feature extraction process.

**(III)** compared our model with the current vascular segmentation best practices. Our ability to efficiently remove speckle-noise while retaining most small retinal vessels in OCTA images with large non-perfusion area offers a distinct advantage.

## Related works

Many deep learning-based automatic segmentations methods have been proposed to address semantic segmentation and medical image issues. Even though these methods may not have been originally designed for OCTA images, they are still effective and considered as typical data segmentation methods that new methods should be evaluated against.

For an example, Olaf et al. [20] developed the famous U-net architecture for cell segmentation. Many researchers have followed Olaf's architecture to develop the U-net++ [21], R2U-Net [22] and Attention U-Net [23] for the segmentation of medical images. DeepLabV3+ [24] and RefineNet [25] are also powerful methods that should be discussed in the semantic segmentation field.

## Proposed methods

We analyzed the vessel segmentation of superficial vascular plexus in OCTA images and proposed an architecture, as illustrated in **Figure 3.** Our architecture was inspired by the original U-net [20] but used a two-stage approach - the backbone stage and the attention stage. The backbone stage focuses on extracting features from OCTA images. The attention stage uses a Gaussian generating layer to fine-tune the results from the backbone stage and enhance vessel detection.

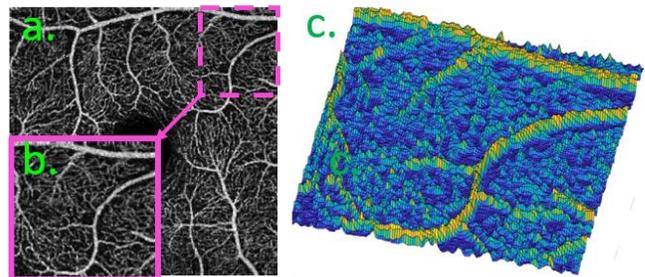

**Figure 4.** Illustration of retinal vessels in OCTA images. **(a)** Original image. **(b)** Magnification of a small patch from the left-upper corner. **(c)** 3D intensity map of (b)

**A. Backbone Stage: feature extraction module.**

Our backbone stage consists of an encoder-decoder architecture to extract features and generate segmentation images. It uses a residual block mechanism [26] to avoid gradient vanishing. It could pass low-level features to high-level layers during backpropagation. The decoder structure adopts a pair parallel network for complex tasks and serves as the symmetric deconvolution mechanism to reconstruct the prediction image.

**B. Attention Stage: Gaussian attention module**

The backbone stage performs a pixel to pixel vessel segmentation but cannot recover continuous details of the shapes of small vessels. To address this problem, we simulated blood vessel distributions using a Gaussian generator. The idea came from observing a blood vessel's distribution, as depicted in **Fig 4 (c)**. The image in Fig 4(c) was made by illustrating 2D vessel images into Gaussian hills using Matlab. We extracted features from the backbone stage to generate cubic features – B(Batch size) × H(Height) × W(Width) × C(Channel). The number of channels (C) in this cubic feature is 6. We then added $confidence$, $\mu_x, \mu_y, \sigma_x, \sigma_y, \rho$ to simulate the bivariate Gaussian distribution as follows (1):

$$f(x,y) = confidence \times \frac{1}{2\pi\sigma_x\sigma_y\sqrt{1-\rho^2}} \times \exp\left(\frac{1}{2(1-\rho^2)}\left[\frac{(x-\mu_x)^2}{\sigma_x^2} + \frac{(y-\mu_y)^2}{\sigma_y^2} - \frac{2\rho(x-\mu_x)(y-\mu_y)}{\sigma_x\sigma_y}\right]\right) \dots (1)$$

$x$ and $y$ are the coordinates of a Gaussian feature map, $confidence \in [0,1]$, $\sigma_x, \sigma_y > 0$ and $\rho > 0$. Our optimizer is an Adam optimizer with a learning rate of 0.00001, batch size of 2 and cross entropy loss function (CE).

Our framework aims to better define vessel morphology through an attention mechanism, identifying and removing noise, and multiplies the backbone stage output to enhance clarity.

## Experiment

**A. Database**

We enrolled OCTA images of 60 eyes from Chang Gung Memorial Hospital, Keelung, Taiwan. The diagnosis were: normal fundus in 11 eyes, retinal vein occlusion in 11 eyes, and diabetic retinopathy of different severities in 38 eyes. Various sizes of non-perfusion area could be found in the diseased eyes. The OCTA images were superficial vascular plexus images with 304 × 304 pixels in size and covered a 3 × 3 mm$^2$ foveal centered area in the posterior pole of fundus. Our database is comprised of the 60 raw OCTA images and the meticulously annotated ground truth of retinal vessels.

Our database was split into 30 training images and 30 images for testing. We adopted four modalities for data augmentation during our training. The modalities include: (i) adjusting brightness by adding random pixel values between -20 to 20; (ii) adding Gaussian noise values with mean = 0 and standard deviations between 0 to 20 or an uniform noise level within the range of -20 to 20; (iii) horizontal and vertical flip; (iv) padding to increase the size of the original image by up to 25% and allowing for the random cropping of images to specified sizes.

**B. Evaluation**

We adopted nine evaluation metrics: (i) Accuracy, (ii) Precision, (iii) Recall (sensitivity), (iv) specificity (v) F1 score (Dice Similarity Coefficient), (vi) AUC, (vii) False Discovery Rate (FDR), (viii) G-means [27] and (ix) Kappa score. The metrics were computed from the number of true-positive, true-negative, false-positive, and false-negative cases (denoted by TP, TN, FP, and FN, respectively):

(i) $\text{Accuracy} = \frac{TP+TN}{TP+TN+FP+FN}$

(ii) $\text{Precision} = \frac{TP}{TP+FP}$

(iii) $\text{Recall (sensitivity)} = \frac{TP}{TP+FN}$

(iv) $\text{Specificity} = \frac{TN}{TN+FP}$

(v) $\text{F1 score} = 2\frac{Recall \cdot Precision}{Recall + precision}$

(vi) $\text{AUC} = \frac{1}{2}\left(\frac{TP}{TP+FN} + \frac{TN}{TN+FP}\right)$

(vii) $\text{FDR} = \frac{FP}{TP+FP}$

(viii) $G - means = \sqrt{sensitivity * Specificity}$

(ix) $Kappa\ score = \frac{Accuracy - pe}{1 - pe}$

$$pe = \frac{((TP+FN)(TP+FP)+(TN+FP)(TN+FN))}{((TP+TN+FP+FN)^2)} \dots (2)$$

**Table 1.** Performance of related methods and ours

| Methods | Acc. | Pre. | Rec. | Spec. | F1 | AUC | FDR | Gme. | Kap. |
|---|---|---|---|---|---|---|---|---|---|
| Deeplabv3+[24] | 0.803 | 0.758 | 0.569 | 0.914 | 0.640 | 0.742 | 0.241 | 0.713 | 0.508 |
| Refinenet [25] | 0.752 | 0.711 | 0.417 | 0.912 | 0.522 | 0.665 | 0.288 | 0.615 | 0.366 |
| U-net [20] | 0.864 | 0.864 | 0.684 | 0.945 | 0.762 | 0.815 | 0.135 | 0.781 | 0.639 |
| R2U-net [22] | 0.854 | 0.844 | 0.675 | 0.933 | 0.745 | 0.804 | 0.155 | 0.791 | 0.641 |
| Attunet [23] | 0.86 | 0.871 | 0.661 | 0.949 | 0.749 | 0.805 | 0.128 | 0.791 | 0.651 |
| U-net++ [21] | 0.857 | **0.874** | 0.646 | **0.951** | 0.739 | 0.798 | **0.125** | 0.781 | 0.639 |
| Ours | **0.865** | 0.816 | **0.755** | 0.917 | **0.781** | **0.836** | 0.183 | **0.831** | **0.681** |

**C. Validation of Segmentation**

Our critical criterion values to quantify our results are: Accuracy, Recall, F1 score, and Kappa score. We compared our model to several current best practice segmentation methods, including six deep learning methods: U-net [20], U-net++ (Unet_nested) [21], R2U-net [22], Attunet [23], Deeplabv3+ [24] and Refinenet [25]. All hyper-parameters in these models were tuned to achieve the best performance. We evaluated them using our database on superficial retinal vascular plexus, and the results are in **Table1**. **Table 1** shows that our segmentation had better performance in the majority of evaluation metrics. Some semantic segmentation methods [24-25] are not suitable for detailed vessel segmentation. Other medical segmentation methods [20-23] could have higher precision score but lower recall values. Nevertheless, our method was superior in all critical metric values, probably because it can better describe the vessel positions using Gaussian attention mechanism.

**Figure 5** illustrates the comparison of output images from our SVS-net and other models. Our SVS-net not only effectively removed the speckle noise, but also showed good ability in preserving small vessels. This may be attributed to the fact that other models only emphasized whether prediction image match the ground truth, our SVS-net's had an additional attention mechanism that can be helpful in reconstructing the morphology of vessels.

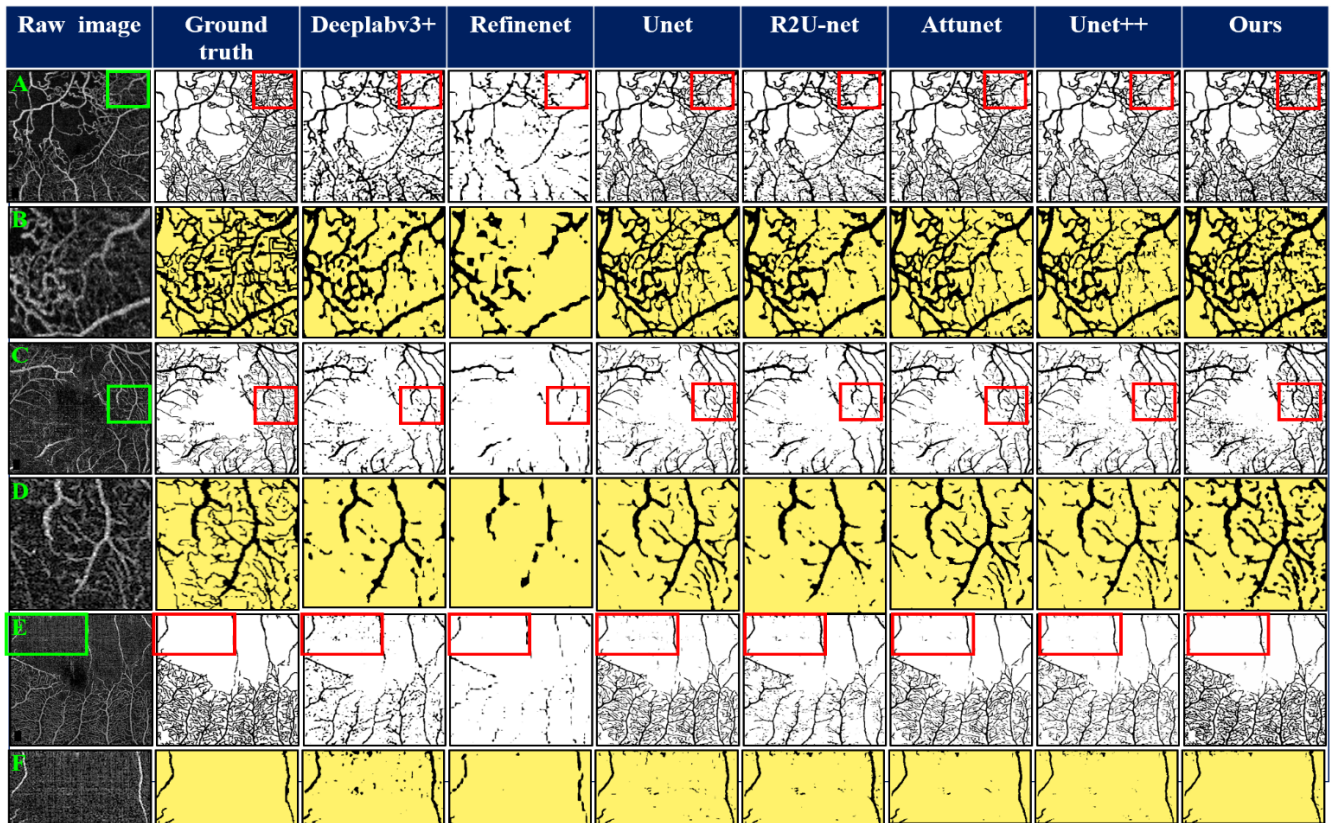

**Figure 5.** Vessel segmentation illustration. (A)(C)(E) are the original raw images; the patches within the rectangular boxes are magnified and shown in (B)(D)(F), respectively.

## Discussions

Using global and local thresholding [4][5] to automatically convert grayscale OCTA images into binary images are acceptable methods in quantitative vascular analyses. In addition, there has been an abundance of research proposing supervised segmentation methods images using manual-annotation ground truth to binarize the OCTA images automatically. However, diffuse speckle noise over a large non-perfusion area may pose a challenge in vessel segmentation. The tradeoff of removing these diffuse speckle noise is usually that small vessel details may also be removed at the same time. The above algorithms have not been validated for their performance in such condition. Furthermore, due to the vessels' complex distribution and ill-defined shape in deep layer vessels, it is difficult for experts to label the ground truth precisely for deep vascular plexus OCTA images. Current OCTA image segmentation best practices do not train their models using these deep layer images.

Therefore, our segmentation model's inference ability is particularly exciting. In **Figure 6,** its ability to retain small vessel details perform well not only in the superficial vascular plexus, but also in the deep vascular plexus. The images produced from our model are similar to manual-annotations with noise removal. We can observe that other pixel to pixel labeling methods sacrificed tiny vessels. Although we did not have the corresponding ground truth data to evaluate our performance on crucial metrics, the advantage of preserving tiny vessels in our model could be demonstrated in some representative areas (red rectangle boxes) in this figure.

## Conclusions

This paper proposed our SVS-net architecture - an novel segmentation model that covers both pixel to pixel segmentation and noise reduction in OCTA images. Our method's effectiveness and practicality are supported by the ground truth established from expertly manually annotated images. We also evaluated our model against the current best practice deep learning methods, including semantic segmentation field and medical segmentation field. We have demonstrated that our model can balance the need to retaining small vessels and the need to remove noise in OCTA images with various sizes of non-perfusion area.

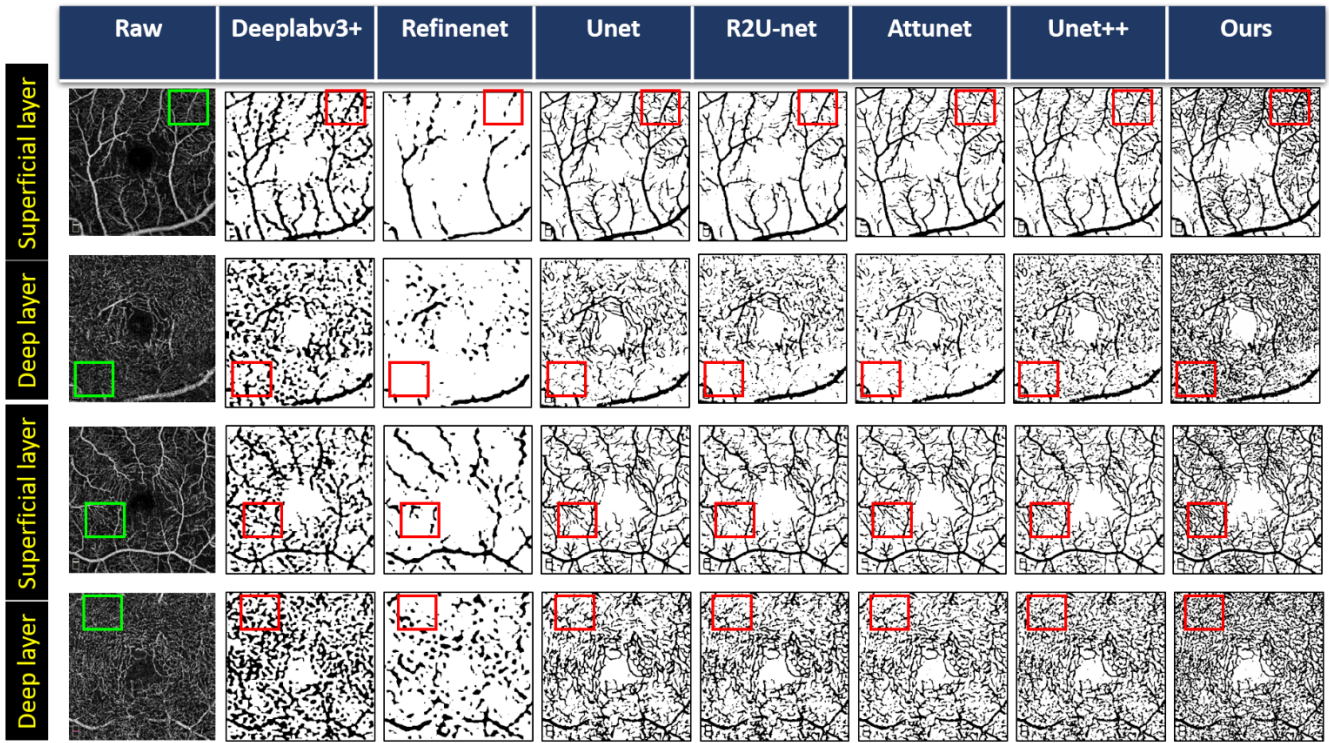

**Figure 6.** Comparison of model capabilities for the OCTA images of the superficial layer and the deep layer.